%% file: article-mine.tex
\documentclass[twoside,twocolumn]{IEEEtran}
\usepackage[T1]{fontenc}
\usepackage[ps,matrix,arrow,curve]{xypic}
\usepackage{amsfonts,amssymb,stmaryrd,amsmath}
\usepackage{epsfig}
\usepackage{url}
\usepackage{galois}
\usepackage{color}

\xyoption{dvips}

\newcommand{\mytop}[1]{{\overline{#1}}}
\newcommand{\itop}{{\,\bar{\imath}}}
\newcommand{\jtop}{{\bar{\jmath}}}

\newcommand{\ktop}{{\bar{k}}}

\newcommand{\mydef}[1]{{\em #1}}

\newenvironment{mytheoreml}
      {\vspace*{0.2cm}\begin{theorem}~\par\begin{enumerate}}
      {\end{enumerate}\hfill$\Box$\end{theorem}}

\newenvironment{mypropertyl}
      {\vspace*{0.2cm}\begin{property}~\par\begin{enumerate}}
      {\end{enumerate}\hfill$\Box$\end{property}}

\newenvironment{mydefinitionl}
      {\vspace*{0.2cm}\begin{definition}~\par\begin{enumerate}}
      {\end{enumerate}\hfill$\Box$\end{definition}}

\newenvironment{mydefinition}
      {\vspace*{0.2cm}\begin{definition}}
      {\par\hfill$\Box$\end{definition}}

\newenvironment{mytheorem}
      {\vspace*{0.2cm}\begin{theorem}}
      {\par\hfill$\Box$\end{theorem}}

\newenvironment{myremark}
      {\vspace*{0.2cm}\begin{remark}}
      {\end{remark}}

\newcommand{\deltaeq}{\;\stackrel{\triangle}{=}\;}
\newcommand{\deltaiff}{\stackrel{\triangle}{\iff}}

\newtheorem{theorem}{Theorem}
\newtheorem{property}{Property}
\newtheorem{definition}{Definition}
\newtheorem{remark}{Remark}

\renewcommand{\vec}[1]{{{\mathbf #1}}}

\begin{document}

% Title
%%%%%%%

\title{The Octagon Abstract Domain}

\author{Antoine Min\'e\\~\\
\small
{\it \'Ecole Normale Sup\'erieure de Paris, France},\\
\url{mine@di.ens.fr},\\
\url{http://www.di.ens.fr/~mine}}

\maketitle

% Abstract
%%%%%%%%%%

\begin{abstract}
This article presents a new numerical abstract domain for static analysis
by abstract interpretation. It extends a former
numerical abstract domain based on Difference-Bound Matrices and allows us to 
represent invariants of the form $(\pm x \pm y \leq c)$, where $x$ and $y$ are
program variables and $c$ is a real constant.

We focus on giving an efficient representation based
on Difference-Bound Matrices---${\mathcal{O}}(n^2)$ memory cost,
where $n$ is the number of variables---and graph-based algorithms
for all common abstract operators---${\mathcal{O}}(n^3)$ time cost.
This includes a normal form algorithm to test equivalence of representation
and a widening operator to compute least fixpoint approximations.
\end{abstract}


% Article
%%%%%%%%%

\section{Introduction}

This article presents practical algorithms to represent and manipulate
invariants of the form $(\pm x \pm y \leq c)$, where $x$ and $y$ are 
numerical variables and $c$ is a numeric constant.
It extends the analysis we previously proposed in our PADO-II article 
\cite{pado2}.
Sets described by such invariants are special kind of polyhedra
called \mydef{octagons} because they feature at most eight edges in dimension 
2 (Figure  \ref{domains}).
Using abstract interpretation, this allows discovering automatically
common errors, such as division by zero, out-of-bound array access
or deadlock, and more generally to prove safety properties for programs.

Our method works well for reals and rationals.
Integer variables can be assumed, in the analysis, to be real in order
to find approximate but safe invariants.

{\bf Example.}
The very simple program described in Figure \ref{sampleprog} simulates
$M$ one-dimensional random walks of $m$ steps and stores the
hits in the array $tab$.
Assertions in curly braces are discovered automatically by a
simple static analysis using our octagonal abstract domain.
Thanks to the invariants discovered, we have the guarantee that the
program does not perform out-of-bound array access at lines 2 and 10.
The difficult point in this example is the fact that the bounds
of the array $tab$ are not known at the time of the analysis; thus,
they must be treated symbolically.

\begin{figure}
\begin{center}
\fbox{
\textsf{
\hspace*{-0.5cm}
\begin{tabular}{rl}
1&{\bf int} $tab[-m\ldots m]$;\\
2&{\bf for} $i=-m$ {\bf to} $m$\quad$tab[i]=0$;\quad$\{-m\leq i\leq m\}$\\
3&{\bf for} $j=1$ {\bf to} $M$ {\bf do}\\
4&\quad{\bf int} $a=0$;\\
5&\quad{\bf for} $i=1$ {\bf to} $m$\\
6&\quad\quad $\{\;1\leq i\leq m;\; -i+1\leq a\leq i-1\;\}$\\
7&\quad\quad {\bf if} $\mbox{rand(2)}=0$\\
8&\quad\quad\quad\quad {\bf then} $a=a+1$;\quad$\{\;-i+1\leq a\leq i\;\}$\\
9&\quad\quad\quad\quad {\bf else} $a=a-1$;\quad$\{\;-i\leq a\leq i-1\;\}$\\
10&\quad $tab[a]=tab[a]+1$;\quad$\{\;-m\leq a\leq m\;\}$\\
11&{\bf done};\\
\end{tabular}}}
\end{center}
\caption{Simulation of a random walk. The assertions in curly brackets
$\{\ldots\}$ are
discovered automatically and prove that this program does not perform
index out of bound error when accessing the array $tab$.}
\label{sampleprog}
\end{figure}

For the sake of brevity, we omit proofs of theorems in this article.
The complete proof for all theorems can be found in the author's
Master thesis \cite{dea}.

\section{Previous Work}

\subsection{Numerical Abstract Domains.}
Static analysis has developed a successful methodology, based on the
abstract interpretation framework---see Cousot and Cousot's
POPL'77 paper \cite{ai}---to build analyzers that discover
invariants automatically:
all we need is an \mydef{abstract domain}, which is
a practical representation of the invariants we want
to study, together with a fixed set of operators and transfer
functions (union, intersection, widening, assignment, guard, etc.) as 
described in Cousot and Cousot's POPL'79 article \cite{semdesign}.

\begin{figure}
\begin{center}
\fbox{\begin{tabular}{ccc}
\epsfig{file=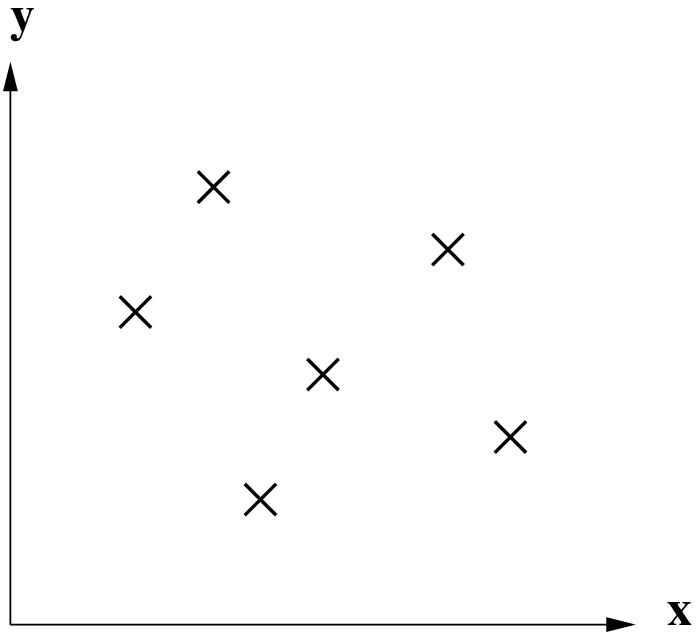,width=3.7cm}&\quad&
\epsfig{file=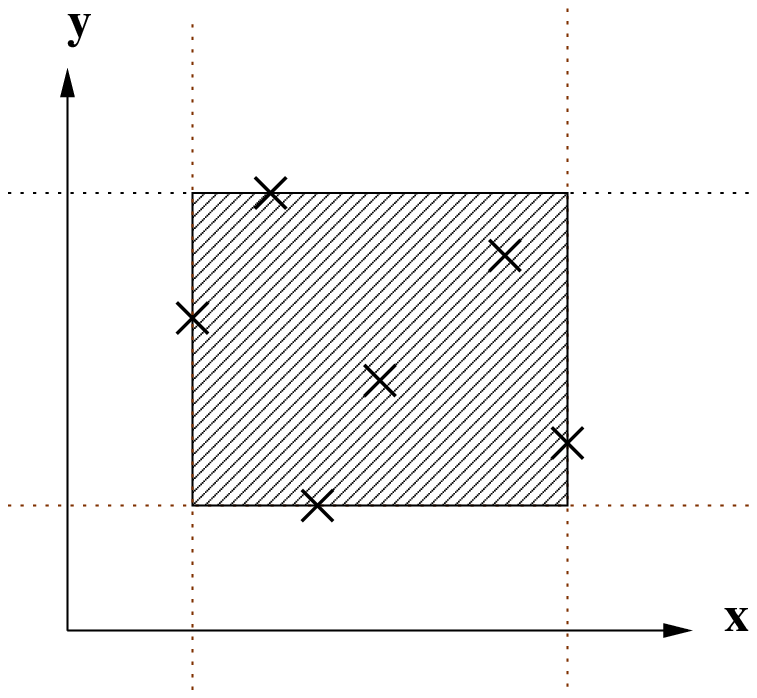,width=3.7cm}\\
(a)&&(b)\\\\
\epsfig{file=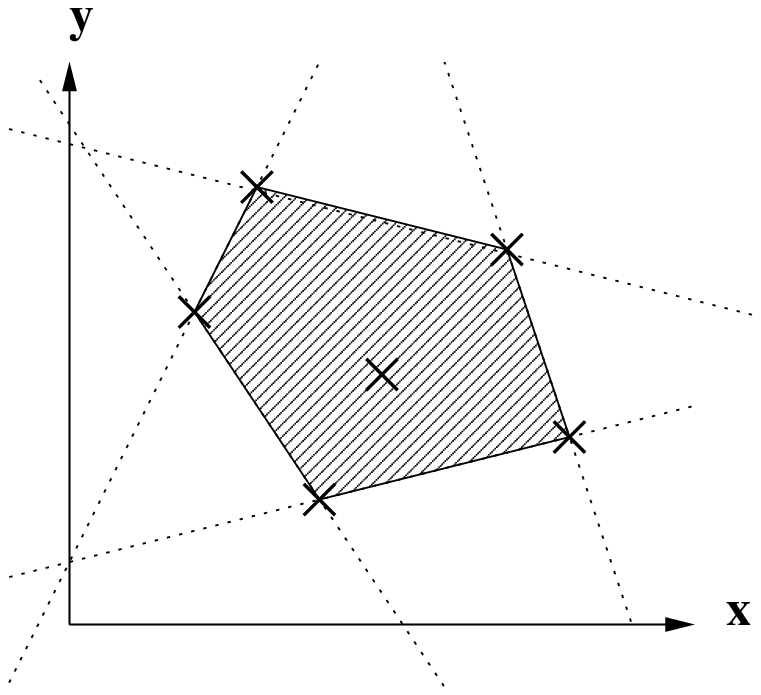,width=3.7cm}&\quad&
\epsfig{file=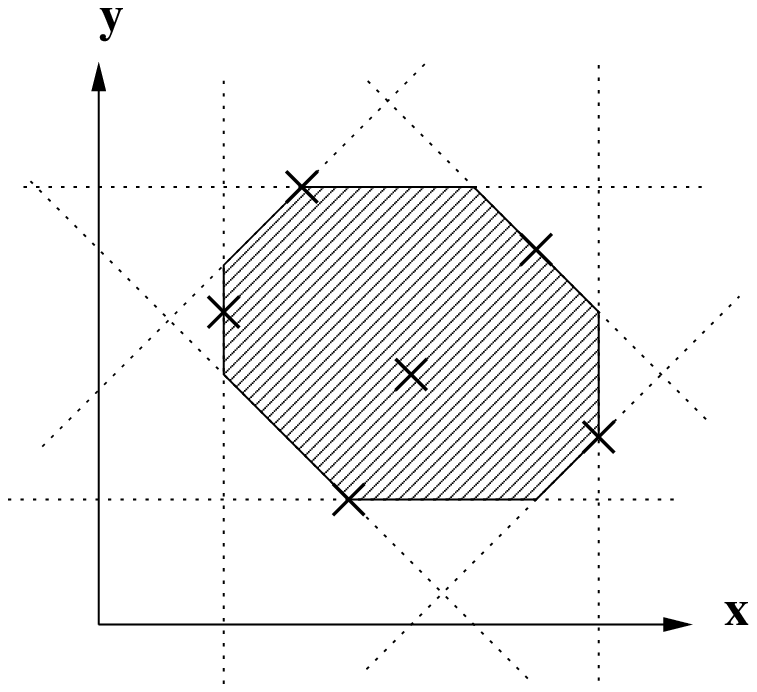,width=3.7cm}\\
(c)&&(d)
\end{tabular}}
\end{center}
\caption{A set of points (a), and its best approximation in the
interval (b), polyhedron (c), and octagon (d) abstract domains.}
\label{domains}
\end{figure}

There exists many \mydef{numerical abstract domains}.
The most used are the lattice of \mydef{intervals} 
(described in Cousot and Cousot's
ISOP'76 article \cite{interv}) and the lattice of \mydef{polyhedra}
(described in Cousot and Halbwachs's POPL'78 article
\cite{poly}).
They represent, respectively, invariants of the form
$(v\in[c_1;c_2])$ and $(\alpha_1v_1+\cdots+\alpha_nv_n\leq c)$, where
$v,v_1,\ldots,v_n$ are program variables and $c, c_1, c_2,\alpha_1,\ldots,
\alpha_n$ are constants.
Whereas the interval analysis is very efficient---linear memory and
time cost---but not very precise, the
polyhedron analysis is much more precise (Figure \ref{domains})
but has a huge memory cost---in practice,
it is exponential in the number of variables.

Remark that the correctness of the program in Figure \ref{sampleprog}
depends on the discovery of invariants of the form $(a\in[-m,m])$ where
$m$ must not be treated as a constant, but 
{\it as a variable}---its value is not known at
analysis time.
Thus, this example is beyond the scope of interval analysis.
It can be solved, of course, using polyhedron analysis.

\subsection{Difference-Bound Matrices.}
Several satisfiability algorithms for set of constraints involving only two
variables per constraint have been proposed in order to solve
{\em Constraint Logic Programming (CLP)} problems.
Pratt analyses, in \cite{pratt}, the simple case of constraints of the form
$(x-y\leq c)$ and $(\pm x\leq c)$ which he called {\em separation theory}.
Shostak then extends, in \cite{shostak}, this to a 
{\em loop residue algorithm} for the case $(\alpha x+\beta y\leq c)$.
However, the algorithm is complete only for reals, not for integers.
Recently, Harvey and Stuckey proposed, in their ACSC'97 article \cite{utvpi},
a complete algorithm, inspired from \cite{shostak},
for integer constraints of the form $(\pm x\pm y\leq c)$.

Unlike CLP, when analyzing programs, we are not only interested in testing the
satisfiability of constraint sets, we also need to manipulate them
and apply operators that mimic the one used to define the semantics of programs
(assignments, tests, control flow junctions, loops, etc.).

The {\em model-checking} community has developed a practical
representation, called {\em Difference-Bound Matrices (DBMs)},
for constraints of the form $(x-y\leq c)$ and ($\pm x\leq c)$,
together with many operators, in order to model-check 
{\em timed automata} (see Yovine's ES'98 article \cite{DBM2} and
Larsen, Larsson, Pettersson, and Yi's RTSS'97 article \cite{DBM}).
These operators are tied to model checking and do not meet the
abstract interpretation needs.
This problem was addressed in our PADO-II article
\cite{pado2} and in
Shaham, Kolodner, and Sagiv's CC2000 article \cite{mooly} which propose
abstract domains based on DBMs, featuring widenings and transfer
functions adapted to real-live programming languages.
All these works are based on the concept of {\em shortest-path closure}
already present in Pratt's article 
\cite{pratt} as the base of the satisfiability
algorithm for constraints of the form $(x-y\leq c)$.
The closure also leads to a normal form that allows easy
equality and inclusion testing.
Good understanding of the interactions between closure and the other operators
is needed to ensure the best precision possible
and termination of the analysis.
These interactions are described in our PADO-II article \cite{pado2}.

Again, proof of the correctness of the program in Figure \ref{sampleprog}
is beyond the scope of the DBM-based abstract domains presented in 
\cite{pado2,mooly} because the invariant $(-a-m\leq 0)$ we need does not
match $(x-y\leq c)$.

\subsection{Our Contribution.}
Our goal is to propose a numerical abstract domain that is between, in term
of expressiveness and cost, the interval and the polyhedron domains.
The set of invariants we discover can be seen as special cases of
linear inequalities; but the underlying algorithmic is very different from
the one used in the polyhedron domain \cite{poly}, and much more efficient.

In this article, we show that DBMs can be extended to describe invariants
of the form $(\pm x \pm y \leq c)$.
We build a new numerical abstract domain, called the
\mydef{octagon abstract domain}, extending the abstract domain
we presented in our PADO-II article \cite{pado2} and
detail algorithms implementing all
operators needed for abstract interpretation.
Most algorithms are adapted from \cite{pado2} but some are much more complex.
In particular, the closure algorithm is replaced by a
{\em strong closure} algorithm.

It is very important to understand that an abstract domain is only a brick
in the design of a static analyzer.
For the sake of simplicity, this paper presents an application of our domain
on a simple forward analysis of a toy programming
language.
However, one could imagine to {\it plug} this domain in various
analyses, such as Bourdoncle's {\sc Syntox} analyzer \cite{syntox},
Deutsch's pointer analysis \cite{deutsch}, 
Dor, Rodeh, and Sagiv's string cleanness checking \cite{nurit}, etc.

Section III recalls the DBM representation for {\em potential constraints}
$(x-y\leq c)$. 
Section IV explains how DBMs can be used to represent a wider range of 
constraints:  interval constraints $(\pm x\leq c)$, and sum 
constraints $(\pm x \pm y\leq c)$.
We then stick to this last extension, as it is the core contribution of this
article, and discuss in Section V about normal form and in 
Section VI about operators and transfer functions.
Section VII builds two lattice structures using these operators. 
Section VIII presents some practical results and gives some ideas for 
improvement.

\section{Difference-Bound Matrices}

In this section, we recall some definitions and simple facts about 
Difference-Bound Matrices (DBMs) and their use in order to represent sets 
of invariants of the form $(x-y\leq c)$.
DBMs are described in \cite{DBM,DBM2} from a model-checking point of view and
in \cite{pado2} for abstract interpretation use.

\medskip

Let ${\mathcal{V}}=\{v_0,\ldots,v_{N-1}\}$ be a finite set of variables with 
value in a numerical set ${\mathbb{I}}$ (which can be ${\mathbb{Z}}$,
${\mathbb{Q}}$ or ${\mathbb{R}}$).
We extend ${\mathbb{I}}$ to $\overline{{\mathbb{I}}}$
by adding the $+\infty$ element; the standard operations
$\leq$, $=$, $+$, $\min$ and $\max$ are extended to $\overline{{\mathbb{I}}}$
as usual.

\subsection{Potential Constraints, DBMs.}

A \mydef{potential constraint over ${\mathcal{V}}$} is a constraint of 
the form $(v_i-v_j\leq c)$, with $v_i,v_j\in{\mathcal{V}}$ and $c\in{\mathbb{I}}$.
Let $C$ be a set of potential constraint over ${\mathcal{V}}$.
We suppose, without loss of generality, that
there do not exist two 
constraints $(v_i-v_j\leq c)$ and $(v_i-v_j\leq d)$ in $C$ 
with $c\neq d$. Then, $C$ can be represented
uniquely by a $N\times N$ matrix $\vec{m}$ with elements in 
$\overline{{\mathbb{I}}}$ :

$$
\vec{m}_{ij}\deltaeq\left\{\begin{array}{ll}
c&\mbox{if $(v_j-v_i\leq c)\in C$},\\
+\infty\quad&\mbox{elsewhere}\enspace.
\end{array}\right.
$$

$\vec{m}$ is called a \mydef{Difference-Bound Matrix (DBM)}.

\subsection{Potential Graph.}

It is convenient to consider $\vec{m}$ as the adjacency matrix of a
weighted graph ${\mathcal{G}}(\vec{m})=\{{\mathcal{V}},{\mathcal{A}},w\}$, 
called its \mydef{potential graph}, and defined by:
$$
\begin{array}{ll}
\begin{array}{l}
{\mathcal{A}}\subseteq{\mathcal{V}}\times{\mathcal{V}},
\\
{\mathcal{A}}\deltaeq\{(v_i,v_j)\;|\;\vec{m}_{ij}<+\infty\},
\end{array}
&
\begin{array}{l}
w\in{\mathcal{A}}\mapsto{\mathbb{I}},
\\
w((v_i,v_j))\deltaeq\vec{m}_{ij}\enspace.
\end{array}
\end{array}
$$

We will denote by 
$\langle i_1,\ldots,i_k \rangle$ a {\em finite} set of
nodes representing a \mydef{path} from node $v_{i_1}$ to node $v_{i_k}$ 
in ${\mathcal{G}}(\vec{m})$.
A \mydef{cycle} is a path such that $i_1=i_k$.

\subsection{$\trianglelefteqslant$ Order.}

The $\leq$ order on $\overline{{\mathbb{I}}}$ induces a point-wise 
partial order $\trianglelefteqslant$ on the set of DBMs:
$$
\vec{m}\trianglelefteqslant\vec{n}\deltaiff \forall i,j,\; 
\vec{m}_{ij}\leq\vec{n}_{ij}\enspace.
$$
The corresponding equality relation is simply the matrix equality $=$.

\subsection{${\mathcal{V}}$-domain.}

Given a DBM $\vec{m}$, the subset of ${\mathcal{V}}\mapsto{\mathbb{I}}$ 
(which will be often assimilated to a subset of ${\mathbb{I}}^N$)
verifying the constraints $\forall i,j,\;v_j-v_i\leq \vec{m}_{ij}$
will be denoted by ${\mathcal{D}}(\vec{m})$ and
called $\vec{m}$'s \mydef{${\mathcal{V}}$-domain}:
$$
{\mathcal{D}}(\vec{m})\deltaeq\{(s_0,\ldots,s_{N-1})\in{\mathbb{I}}^N\;|\;
\forall i,j,\;s_j-s_i\leq \vec{m}_{ij}\}\enspace.
$$

\begin{figure}
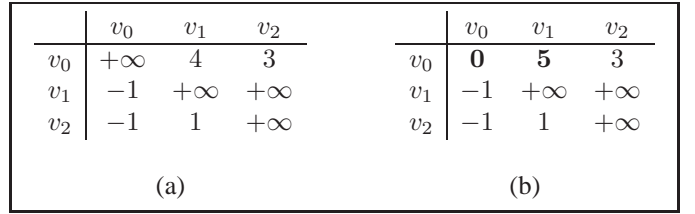

\begin{center}
\fbox{\begin{tabular}{ccc}
$
\begin{array}{c|ccc}
& v_0 & v_1 & v_2\\
\hline
v_0 & +\infty & 4 & 3 \\
v_1 & -1 & +\infty & +\infty\\
v_2 & -1 & 1 & +\infty\\
\end{array}
$
&\quad\quad&
$
\begin{array}{c|ccc}
& v_0 & v_1 & v_2\\
\hline
v_0 & {\bf 0} & {\bf 5} & 3 \\
v_1 & -1 & +\infty & +\infty\\
v_2 & -1 & 1 & +\infty\\
\end{array}
$
\\\\
(a)&&(b)
\end{tabular}}
\end{center}
\caption{Two different DBMs with the same 
${\mathcal{V}}$-domain. Remark that (a) and (b) are not even comparable 
with respect to $\trianglelefteqslant$.}
\label{uniqness}
\end{figure}

By extension, we will call \mydef{${\mathcal{V}}$-domain} any subset of 
${\mathcal{V}}\mapsto{\mathbb{I}}$ which is the ${\mathcal{V}}$-domain of some DBM.
\begin{myremark}
We have $\vec{m}\trianglelefteqslant\vec{n}\Longrightarrow
{\mathcal{D}}(\vec{m})\subseteq{\mathcal{D}}(\vec{n})$, but the converse is false.
As a consequence, representation of ${\mathcal{V}}$-domains is not unique 
and we can have
${\mathcal{D}}(\vec{m})={\mathcal{D}}(\vec{n})$ but $\vec{m}\neq\vec{n}$
(Figure  \ref{uniqness}).
\end{myremark}

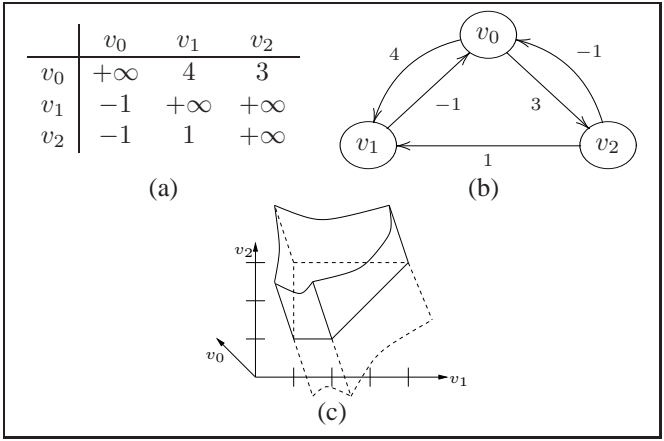
\begin{figure}
\begin{center}
\fbox{\begin{tabular}{cc}
$
\begin{array}{c|ccc}
& v_0 & v_1 & v_2\\
\hline
v_0 & +\infty & 4 & 3 \\
v_1 & -1 & +\infty & +\infty\\
v_2 & -1 & 1 & +\infty\\
\end{array}
$
&
\raisebox{0.8cm}{
\xymatrix{&
*++[o][F-]{v_0} \ar@/_1pc/[ld]_4 \ar[rd]_3 & \\
*++[o][F-]{v_1} \ar[ur]_{-1} & & 
*++[o][F-]{v_2} \ar@/_1pc/[ul]_{-1} \ar[ll]^1
}}
\\
(a)&(b)\\
\multicolumn{2}{c}{\raisebox{-1.5cm}{\input{vdomain.pstex_t}}}\\
\multicolumn{2}{c}{(c)}
\end{tabular}}
\end{center}
\caption{A DBM (a), its potential graph (b) and its 
${\mathcal{V}}$-domain (c).}
\label{dbm}
\end{figure}

\section{Extending Difference-Bound Matrices}

Discovering invariants of the single potential form $(x-y\leq c)$
is not very interesting; however DBMs can be used to represent broader
constraint forms.
In this section,
we first present briefly how to add interval constraints $(\pm x\leq c)$.
This extension is not new: \cite{DBM,pado2} use it instead of pure DBM.
We then present our new extension allowing representation of the more
general constraints $(\pm x \pm y \leq c)$.

\subsection{Representing intervals.}
Given a finite set of variables ${\mathcal{V}}^0=\{v_0,\ldots,v_{N-1}\}$,
in order to represent constraints of the form 
$(v_i-v_j\leq c)$ and $(\pm v_i\leq c)$, we simply add to ${\mathcal{V}}^0$ a 
special variable, named ${\bf 0}$, which is supposed to be always equal to
0.
Constraints of the form ($v_i\leq c)$ and $(v_j\geq d)$ can then be rewritten
as $(v_i-{\bf 0}\leq c)$ and $({\bf 0}-v_j\leq -d)$,
which are indeed potential constraints
over the set ${\mathcal{V}}=\{{\bf 0},v_0,\ldots,v_{N-1}\}$.

We will use a $0$ superscript to denote that a DBM over 
${\mathcal{V}}$ represents a set of extended constraints over ${\mathcal{V}}^0$.
Given such a DBM $\vec{m}^0$,  we
will not be interested in its ${\mathcal{V}}$-domain, ${\mathcal{D}}(\vec{m}^0)$,
which is a subset of ${\mathcal{V}}\mapsto{\mathbb{I}}$, 
but in its \mydef{${\mathcal{V}}^0$-domain},
denoted by ${\mathcal{D}}^0(\vec{m}^0)$ and defined by:
$$
{\mathcal{D}}^0(\vec{m}^0)
\deltaeq\left\{
\begin{array}{ll}
(s_0,\ldots,s_{N-1})\in\mathbb{I}^N\;|\;\\
(0,s_0,\ldots,s_{N-1})\in {\mathcal{D}}(\vec{m}^0)
\end{array}\right\}\subseteq{\mathcal{V}}^0\mapsto{\mathbb{I}}\enspace.
$$

We will call \mydef{${\mathcal{V}}^0$-domain} any subset of 
${\mathcal{V}}^0\mapsto{\mathbb{I}}$ which is the ${\mathcal{V}}^0$-domain of
some DBM $\vec{m}^0$.
As before, $\vec{m}^0\trianglelefteqslant\vec{n}^0\Longrightarrow
{\mathcal{D}}^0(\vec{m}^0)\subseteq{\mathcal{D}}^0(\vec{n}^0)$, but the converse
is false.

\subsection{Representing sums.}

We suppose that ${\mathcal{V}}^+=\{v_0,\ldots,v_{N-1}\}$ is a finite set of 
variables.
The goal of this article is to present a new DBM extension
adapted to represent constraints of the form $(\pm v_i \pm v_j \leq c)$,
with $v_i,v_j\in{\mathcal{V}}^+$ and $c\in{\mathbb{I}}$.

In order to do this, we consider that each variable $v_i$ in ${\mathcal{V}}^+$
comes in two flavors: a positive form $v^+_i$ and a negative form
$v^-_i$. We introduce the set
${\mathcal{V}}=\{\;v^+_0,\;v^-_0,\;\ldots,\;v^+_{N-1},\;v^-_{N-1}\;\}$
and consider DBMs over ${\mathcal{V}}$.
Within a potential constraint, a positive variable $v^+_i$ will be 
interpreted as $+v_i$, and a negative variable $v^-_i$ as $-v_i$; thus
it is possible to represent $(v_i+v_j\leq c)$ by $(v^+_i-v^-_j\leq c)$.
More generally, any set of constraints of the form $(\pm v_i \pm v_j\leq c)$,
with $v_i,v_j\in{\mathcal{V}}^+$ can be represented by a DBM over
${\mathcal{V}}$, following the translation described in Figure  \ref{trans}.
\begin{myremark}
We do not need to add a special variable ${\bf 0}$ to
represent interval constraints as we did before. 
Constraints of the form $(v_i\leq c)$ and $(v_i\geq c)$ can be represented as
$(v^+_i-v^-_i\leq 2c)$ and $(v^-_i-v^+_i\leq -2c)$.
\end{myremark}

\begin{figure}
\begin{center}
$
\begin{array}{|cc|cc|}
\hline
\multicolumn{2}{|c|}{\mbox{constraint over ${\mathcal{V}}^+$}} &
\multicolumn {2}{c|}{\mbox{constraint(s) over ${\mathcal{V}}$}} \\
\hline
v_i-v_j\leq c & (i\neq j) &
v_i^+-v_j^+\leq c, & v_j^--v_i^-\leq c \\

v_i+v_j\leq c & (i\neq j) &
v_i^+-v_j^-\leq c, & v_j^+-v_i^-\leq c \\

-v_i-v_j\leq c & (i\neq j) &
v_j^--v_i^+\leq c, & v_i^--v_j^+\leq c \\

v_i\leq c & & 
\multicolumn{2}{c|}{v_i^+-v_i^-\leq 2c}\\

v_i\geq c & &
\multicolumn{2}{c|}{v_i^--v_i^+\leq -2}\\
\hline
\end{array}
$
\end{center}
\caption{Translation between extended constraints over ${\mathcal{V}}^+$ and 
potential constraints over ${\mathcal{V}}$.}
\label{trans}
\end{figure}

\subsection{Index Notation.}
We will use a $+$ superscript to denote that a DBM over
${\mathcal{V}}$ represents a set of extended constraints over ${\mathcal{V}}^+$.
Such a DBM $\vec{m}^+$ is a $2N\times 2N$ matrix with the following convention:
a row or column index of the form $2i,\;i<N$ corresponds to the variable
$v^+_i$ and an index of the form $2i+1,\;i<N$ corresponds to the variable
$v^-_i$.

We introduce the $\cdot\mapsto\mytop{\cdot}$ operator on indices
defined by $\itop\deltaeq i\oplus 1$---where $\oplus$ is the
{\em bit-wise exclusive or} operator---so that, if $i$ corresponds to 
$v^+_j$, then 
$\itop$ corresponds to $v^-_j$ and if $i$ corresponds to $v^-_j$, then 
$\itop$ corresponds to $v^+_j$.

\subsection{Coherence.}
Figure  \ref{trans} shows that some constraints over ${\mathcal{V}}^+$ can be 
represented by different potential constraints over ${\mathcal{V}}$.
A DBM $\vec{m}^+$ will be said to be \mydef{coherent} if two potential 
constraints 
over ${\mathcal{V}}$ corresponding to the same constraint over ${\mathcal{V}}^+$
are either both represented in $\vec{m}^+$, or both absent.
Thanks to the $\cdot\mapsto\mytop{\cdot}$ operator we introduced, coherence
can be easily characterized:
\begin{mytheorem}
$\vec{m}^+\mbox{ is coherent }\iff 
\forall i,j,\;\vec{m}^+_{ij}=\vec{m}^+_{\jtop\itop}\enspace.$
\end{mytheorem}

In the following, DBMs with a $+$ superscript 
will be assumed to be coherent.

\subsection{${\mathcal{V}}^+$-domain.}

As for the simple interval extension, the ${\mathcal{V}}$-domain 
of a DBM $\vec{m}^+$ is not of interest:
we need to get back in ${\mathcal{V}}^+\mapsto{\mathbb{I}}$ and take into account
the fact that variables in ${\mathcal{V}}$ are not independent but
related by $v^+_i=-v^-_i$.
Thus, we define the \mydef{${\mathcal{V}}^+$-domain} of $\vec{m}^+$ and denote by
${\mathcal{D}}^+(\vec{m}^+)$ the set:
$$
{\mathcal{D}}^+(\vec{m}^+)
\deltaeq\left\{
\begin{array}{ll}
(s_0,\ldots,s_{N-1})\in{\mathbb{I}}^N\;|\;\\
(s_0,-s_0,\ldots,s_{N-1},-s_{N-1})\in {\mathcal{D}}(\vec{m}^+)
\end{array}\right\}.
$$

We will call \mydef{octagon} any subset of 
${\mathcal{V}}^+\mapsto{\mathbb{I}}$ which is the ${\mathcal{V}}^+$-domain of
some coherent DBM $\vec{m}^+$.
As before, $\vec{m}^+\trianglelefteqslant\vec{n}^+\Longrightarrow
{\mathcal{D}}^+(\vec{m}^+)\subseteq{\mathcal{D}}^+(\vec{n}^+)$, but the converse
is false.

\section{Emptiness Test and Normal Forms}

We saw in Figure  \ref{uniqness} that two different DBMs can 
have the same ${\mathcal{V}}$-domain.
Fortunately, there exists a normal form for DBMs representing non-empty
octagons.

In this section, we first recall the normal form for classical DBMs $\vec{m}$,
and then show how it can be adapted to DBMs $\vec{m}^+$
representing non-empty octagons.
Unfortunately, our adaptation does not work very well with
integers.

The potential graph interpretation of DBMs will be very helpful to
understand the algorithms presented.

\subsection{Emptiness Test.}

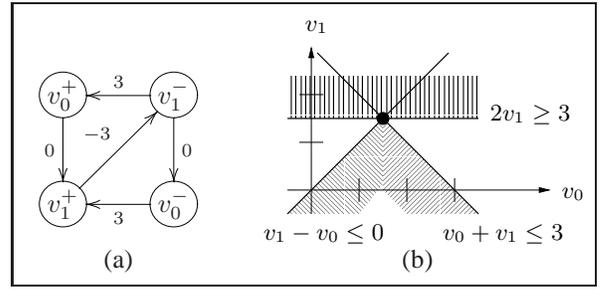
\begin{figure}
\begin{center}
\fbox{\begin{tabular}{ccc}
\xymatrix{
*+[o][F-]{v_0^+} \ar[d]_{0} & *+[o][F-]{v_1^-} \ar[d]^{0} \ar[l]_{3} \\
*+[o][F-]{v_1^+} \ar[ur]^{-3} & *+[o][F-]{v_0^-} \ar[l]^{3} \\
}
&\quad&
\raisebox{-2cm}{\input{emptyZ.pstex_t}}
\\
(a)&&(b)
\end{tabular}}
\end{center}
\caption{A potential graph ${\mathcal{G}}(\vec{m}^+)$
in ${\mathbb{Z}}$ with no strictly negative cycle (a)
and the corresponding ${\mathcal{V}}^+$-domain (b)
${\mathcal{D}}^+(\vec{m}^+)=\{(\frac{3}{2},\frac{3}{2})\}$ which is empty
in ${\mathbb{Z}}^2$.}
\label{emptypb}
\end{figure}

The following graph-oriented theorem allows us to perform emptiness testing
for ${\mathcal{V}}$-domains, ${\mathcal{V}}^0$-domains and octagons:

\begin{mytheoreml}
\label{emptythm}
\item
${\mathcal{D}}(\vec{m})=\emptyset\iff{\mathcal{G}}(\vec{m})$ has
a cycle with a strictly negative weight.
\item ${\mathcal{D}}(\vec{m}^0)=\emptyset\iff
{\mathcal{D}}^0(\vec{m}^0)=\emptyset$.
\item 
If ${\mathbb{I}}\neq{\mathbb{Z}}$, then
${\mathcal{D}}(\vec{m}^+)=\emptyset\iff
{\mathcal{D}}^+(\vec{m}^+)=\emptyset$.

If ${\mathbb{I}}={\mathbb{Z}}$, then
${\mathcal{D}}(\vec{m}^+)=\emptyset\Longrightarrow
{\mathcal{D}}^+(\vec{m}^+)=\emptyset$, but the converse is false
(Figure  \ref{emptypb}).
\end{mytheoreml}

If ${\mathbb{I}}\neq{\mathbb{Z}}$,
in order to check whether the ${\mathcal{V}}^+$-domain of a DBM $\vec{m}^+$ is 
empty, 
we simply have to check for cycles with a strictly negative weight 
in ${\mathcal{G}}(\vec{m}^+)$ using,
for example, the well-known {\em Bellman-Ford algorithm} which runs
in ${\mathcal{O}}(N^3)$ time and is described in 
Cormen, Leiserson and Rivest's classical algorithmic textbook
\cite[\S 25.3]{CLR}.

Figure \ref{emptypb} gives an example where our algorithm fails when dealing
with integers.
Indeed, we have ${\mathcal{D}}(\vec{m}^+)=\{(3+x,3-x,3+y,3-y)\;|\;
\forall x,y\in{\mathbb{Z}}\}$ which is not empty, but all these solutions
over $\{v_0^+,v_0^-,v_1^+,v_1^-\}$ correspond to
the singleton $\{(3/2,3/2)\}$ when we get back to
$\{v_0,v_1\}$, which is not an acceptable solution
in ${\mathbb{Z}}^2$, so
${\mathcal{D}}^+(\vec{m}^+)$ should be empty.
The problem is that a DBM $\vec{m}^+$ with coefficients in
${\mathbb{Z}}$ can represent constraints that use not only integers, but also
half-integers constants---such as $v_1\geq 3/2$ in Figure \ref{emptypb}.

\subsection{Closure.}

Given a DBM $\vec{m}$, the ${\mathcal{V}}$-domain of which is not empty,
${\mathcal{G}}(\vec{m})$ has no strictly negative cycle, so its
{\em shortest-path closure}---or simply {\em closure}---$\vec{m}^*$ 
is well-defined by:
$$
\left\{\begin{array}{ll}
\vec{m}^*_{ii}\deltaeq 0,&\\
\displaystyle
\vec{m}^*_{ij}\deltaeq\min_{\substack{
1\leq M\\ 
\langle i=i_1,i_2,\ldots, i_M=j\rangle\quad}}
\sum_{k=1}^{M-1} \vec{m}_{i_k i_{k+1}}\quad&\mbox{ if }i\neq j\enspace.
\end{array}
\right.
$$

The idea of closure relies on the fact that,
if $\langle i=i_1,i_2,\ldots,i_M=j \rangle$
is a path from $v_i$ to $v_j$, then the constraint
$v_j-v_i \leq \sum_{k=1}^{M-1} \vec{m}_{i_k i_{k+1}}$ can be derived from 
$\vec{m}$ by adding the potential constraints 
$v_{i_{k+1}}-v_{i_k} \leq \vec{m}_{i_k i_{k+1}},\;1\leq k\leq M-1$.
This is an \mydef{implicit} potential constraint as it does not appear 
directly in $\vec{m}$.
In the closure, we replace each potential constraint
$v_j-v_i\leq \vec{m}_{ij}$
by the tightest implicit constraint we can find by summation over
paths of ${\mathcal{G}}(\vec{m})$ if $i\neq j$, or by $0$ 
if $i=j$ ($0$ is indeed the smallest value taken by $v_i-v_i$).

We have the following theorem:
\begin{mytheoreml}
\label{closurethm}
\item $\vec{m}=\vec{m}^*\iff\forall i,j,k,\;
\vec{m}_{ij}\leq \vec{m}_{ik}+\vec{m}_{kj}$ and
$\forall i,\;\vec{m}_{ii}=0$ {\it (Local Definition)}.
\item $\forall i,j,\;$ if $\vec{m}^*_{ij}\neq+\infty$, then
$\exists(s_0,\ldots,s_{N-1})\in{\mathcal{D}}(\vec{m})$
such that $s_j-s_i=\vec{m}^*_{ij}$
{\it (Saturation)}.
\item
$\vec{m}^*=\inf_{\trianglelefteqslant} \{\vec{n}\;|\;{\mathcal{D}}(\vec{n})
={\mathcal{D}}(\vec{m})\}$ {\it (Normal Form)}.
\end{mytheoreml}

Theorem \ref{closurethm}.2 proves that the closure is indeed a normal form.
Theorem \ref{closurethm}.1 leads to a closure algorithm
inspired by the {\em Floyd-Warshall} shortest-path algorithm.
This algorithm is described in Figure  \ref{closurealg} and runs
in ${\mathcal{O}}(N^3)$ time.
Theorem  \ref{closurethm}.2 is crucial to analyze precision of some operators
(such as projection and union).

\begin{myremark}
The closure is also a normal form for DBMs representing
non-empty ${\mathcal{V}}^0$-domains:\\
$(\vec{m}^0)^*=\inf_{\trianglelefteqslant} \{
\vec{n}^0\;|\;{\mathcal{D}}^0(\vec{n}^0)
={\mathcal{D}}^0(\vec{m}^0)\}$.
\end{myremark}

\begin{figure}
\begin{center}
\fbox{\quad
$\begin{array}{c}
\left\{
\begin{array}{lcl}
\vec{m}_0&\deltaeq&\vec{m},\\
\vec{m}_{k+1}&\deltaeq&C_k(\vec{m}_k)\quad\forall k,\;0\leq k<N,\\
\vec{m}^*&\deltaeq&\vec{m}_N,
\end{array}
\right.
\\\\
\mbox{where $C_k$ is defined, $\forall k$, by:}
\\\\
\left\{
\begin{array}{l}
\left[C_k(\vec{n})\right]_{ii}\deltaeq0,\\
\left[C_k(\vec{n})\right]_{ij}\deltaeq
\min(\vec{n}_{ij},\vec{n}_{ik}+\vec{n}_{kj})
\quad\forall i\neq j\enspace.
\end{array}
\right.
\end{array}$
\quad}
\end{center}
\caption{Closure algorithm derived from the {\em Floyd-Warshall} 
shortest-path algorithm.}
\label{closurealg}
\end{figure}

\subsection{Strong Closure.}
\begin{figure}
\begin{center}
\fbox{\begin{tabular}{ccc}
\raisebox{0.7cm}{\xymatrix{
*+[o][F-]{v_0^+} & *+[o][F-]{v_1^-} \ar[d]^{4} \\ 
*+[o][F-]{v_0^-} \ar[u]^2 & *+[o][F-]{v_1^+}
}}
&\quad\quad$\Longrightarrow$\quad\quad&
\raisebox{0.7cm}{\xymatrix{
*+[o][F-]{v_0^+} & *+[o][F-]{v_1^-} \ar[d]^{4} \ar[l]_{3} \\
*+[o][F-]{v_0^-} \ar[u]^2 \ar[r]_{3} & *+[o][F-]{v_1^+}
}}\\
(a)&&(b)
\end{tabular}}
\end{center}
\caption{A DBM (a) and its strong closure (b).
Note that (a) is closed, and that (a) and (b) have the same 
${\mathcal{V}}^+$-domain but not the same ${\mathcal{V}}$-domain.
In (b), we deduced $(v_0+v_1\leq 3)$ from $(2v_0\leq 2)$ and $(2v_1\leq 4)$,
so it is smaller than (a) with respect to $\trianglelefteqslant$.}
\label{closurepb}
\end{figure}
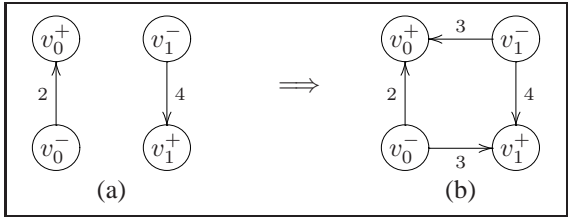

We now focus on finding a normal form for DBMs representing 
non-empty octagons.
The solution presented above does not work because two different
DBMs can have the same ${\mathcal{V}}^+$-domain
but different ${\mathcal{V}}$-domains,
and so the closure $(\vec{m}^+)^*$ of $\vec{m}^+$ is not
the smallest DBM---with respect to the $\trianglelefteqslant$ order---that
represents the octagon ${\mathcal{D}}^+(\vec{m}^+)$.
The problem is that the set of implicit constraints gathered by summation of 
constraints over paths of ${\mathcal{G}}(\vec{m}^+)$ is not sufficient.
Indeed, we would like to deduce $(v_i^+-v_j^-\leq (c+d)/2)$ from
$(v_i^+-v_i^-\leq c)$ and $(v_j^+-v_j^-\leq d$), which is not possible because
the set of edges $\{(v_i^-,v_i^+),(v_j^-,v_j^+)\}$ does not form a path
(Figure  \ref{closurepb}).

Here is a more formal description of a normal form,
called the \mydef{strong closure}, adapted from the closure:

\begin{mydefinition}
\label{strongclosuredef}
$\vec{m}^+$ is {\em strongly closed} if and only if
\begin{itemize}
\item $\vec{m}^+$ is {\it coherent}: 
$\forall i,j,\;\vec{m}^+_{ij}=\vec{m}^+_{\jtop \itop}$;
\item $\vec{m}^+$ is {\it closed}:
$\forall i,\;\vec{m}^+_{ii}=0$ and
$\forall i,j,k,\;
\vec{m}^+_{ij}\leq\vec{m}^+_{ik}+\vec{m}^+_{kj}$;
\item $\forall i,j,\;$ $\vec{m}^+_{ij}\leq (\vec{m}^+_{i \itop}
+\vec{m}^+_{\jtop j})/2$.
\end{itemize}
\end{mydefinition}

From this definition, we derive the \mydef{strong closure algorithm}
$\vec{m}^+\mapsto(\vec{m}^+)^\bullet$
described in Figure  \ref{strongclosurealg}.
The algorithm looks a bit like the closure algorithm of Figure  \ref{closurealg}
and also runs in ${\mathcal{O}}(N^3)$ time.
It uses two auxiliary functions $C^+_k$ and $S^+$.
The $C^+_k$ function looks like the $C_k$ function 
used in the closure algorithm
except it is designed to maintain coherence;
each $C^+_k$ application is a step toward closure.
The $S^+$ function ensures that 
$\forall i,j,\;\left[S^+(\vec{m}^+)\right]_{ij}\leq
(\left[S^+(\vec{m}^+)\right]_{i\itop}+
 \left[S^+(\vec{m}^+)\right]_{\jtop j})/2$
while maintaining coherence.

There is no simple explanation for the complexity of $C^+_k$;
the five terms in the {\em min} 
statement appear naturally when trying to prove 
that, when interleaving $C^+_k$ and $S^+$ steps, what was gained before
will not be destroyed in the next step.

The following theorem holds for ${\mathbb{I}}\neq{\mathbb{Z}}$:

\begin{mytheoreml}
\label{strongclosurethm}
\item $\vec{m}^+=(\vec{m}^+)^{\bullet} \iff \vec{m}^+\mbox{ is strongly closed}$.
\item $\forall i,j,\;$ if $(\vec{m}^+)^\bullet_{ij}\neq+\infty$, then
$\exists(s_0,\ldots,s_{2N-1})\in{\mathcal{D}}(\vec{m}^+)$
such that $\forall k,\; s_{2k}=-s_{2k+1}$ 
and $s_j-s_i=(\vec{m}^+)^\bullet_{ij}$ {\it (Saturation)}.
\item $(\vec{m}^+)^\bullet=
\inf_{\trianglelefteqslant} \{\vec{n}^+\;|\;{\mathcal{D}}^+(\vec{n}^+)
={\mathcal{D}}^+(\vec{m}^+)\}$ {\it (Normal Form)}.
\end{mytheoreml}

This theorem is very similar to Theorem \ref{closurethm}.
It states that, when ${\mathbb{I}}\neq{\mathbb{Z}}$,
the strong closure algorithm
gives a strongly closed DBM (Theorem \ref{strongclosurethm}.1)
which is indeed a normal form (Theorem \ref{strongclosurethm}.3).
The nice saturation property of Theorem \ref{strongclosurethm}.2 
is useful to analyze the projection and union operators.

\begin{figure}
\begin{center}
\fbox{
$\begin{array}{c}
\left\{
\begin{array}{lcl}
\vec{m}^+_0&\deltaeq&\vec{m}^+,\\
\vec{m}^+_{k+1}&\deltaeq&S^+(C^+_{2k}(\vec{m}^+_k))
\quad\forall k,\;0\leq k<N,\\
(\vec{m}^+)^\bullet&\deltaeq&\vec{m}^+_{N},\\
\end{array}
\right.
\\\\
\mbox{where $C^+_k$ is defined, $\forall k$, by:}
\\\\
\left\{
\begin{array}{ll}
\left[C^+_k(\vec{n}^+)\right]_{ii}\deltaeq0,\\ 
\left[C^+_k(\vec{n}^+)\right]_{ij}\deltaeq\min(&
\vec{n}^+_{ij},\;(\vec{n}^+_{ik}+\vec{n}^+_{kj}),\\
&(\vec{n}^+_{i \ktop}+\vec{n}^+_{\ktop j}),\\
&(\vec{n}^+_{ik}+\vec{n}^+_{k \ktop}+\vec{n}^+_{\ktop j}),\\
&(\vec{n}^+_{i \ktop}+\vec{n}^+_{\ktop k}+\vec{n}^+_{k j})\;)
\end{array}
\right.
\\\\
\mbox{and $S^+$ is defined by:}
\\\\
\begin{array}{l}
\left[S^+(\vec{n}^+)\right]_{ij}\deltaeq\min(\;\vec{n}^+_{ij},\;
(\vec{n}^+_{i\itop}+\vec{n}^+_{\jtop j})/2\;)\enspace.
\end{array}
\end{array}$}
\end{center}
\caption{Strong Closure algorithm.}
\label{strongclosurealg}
\end{figure}

%\begin{figure}
%\begin{center}
%\begin{tabular}{ccc}
%\raisebox{0.7cm}{\xymatrix{
%*+[o][F-]{v_0^+} \ar@/^/[r]^{0} & *+[o][F-]{v_0^-} \ar@/^/[l]^{0} \\ 
%*+[o][F-]{v_1^+} \ar[r]_{1} & *+[o][F-]{v_1^-}
%}}
%&\quad\quad$\Longrightarrow$\quad\quad&
%\raisebox{0.7cm}{\xymatrix{
%*+[o][F-]{v_0^+} \ar@/^/[r]^{0} & *+[o][F-]{v_0^-} \ar@/^/[l]^{0} \ar[d]^{0}\\ 
%*+[o][F-]{v_1^+} \ar[r]_{\bf 1} \ar[u]^{0} & *+[o][F-]{v_1^-}
%}}
%\end{tabular}
%\end{center}
%\caption{A DBM on $\overline{{\mathbb{Z}}}$ (a) and its strong closure (b)
%which is not closed.}
%\label{strongclosurepb}
%\end{figure}

\subsection{Discussions about ${\mathbb{Z}}$.}
Classical DBMs and the interval constraint extension work equally
well on reals, rationals and integers.
However, our extension does not handle integers properly.

When ${\mathbb{I}}={\mathbb{Z}}$,
the strong closure algorithm does not lead
to the smallest DBM with the same ${\mathcal{V}}^+$-domain.
For example, knowing that $x$ is an integer,
the constraint $2x\leq 2c$ should be deduced from $2x\leq 2c+1$, which
the strong closure algorithm fails to do.
More formally, Definition  \ref{strongclosuredef} is not sufficient; our 
normal form
should also respect: $\forall i,\;\vec{m}^+_{i\itop}\mbox{ is even}$.
One can imagine to simply add to the strong closure algorithm a 
{\em rounding phase} $R^+$ defined by 
$\left[R^+(\vec{m}^+)\right]_{i\itop}=2\lfloor\vec{m}^+_{i\itop}/2\rfloor$
and $\left[R^+(\vec{m}^+)\right]_{i j}=\vec{m}^+_{ij}$ if $i\neq\jtop$,
but it is tricky to make $R^+$ and $C^+_k$ interact correctly
so we obtain a DBM which is {\em both} closed and rounded.
We were unable, at the time of writing, to design such an algorithm 
and keep a ${\mathcal{O}}(N^3)$ time cost.

This problem was addressed by Harvey and Stuckey in their ACSC'97 article 
\cite{utvpi}.
They propose a satisfiability algorithm mixing closure and tightening steps 
that can be used to test emptiness and build the normal form 
$(\vec{m}^+)^\bullet=
\inf_{\trianglelefteqslant} \{\vec{n}^+\;|\;{\mathcal{D}}^+(\vec{n}^+)
={\mathcal{D}}^+(\vec{m}^+)\}$ we need.
Unfortunately, this algorithm has a ${\mathcal{O}}(N^4)$ time cost
in the worst case.
This algorithm has the advantage of being incremental---${\mathcal{O}}(N^2)$ 
time cost per constraint changed in the DBM---which 
is useful for CLP problems but does not seem interesting in
static analysis because many operators are point-wise and change all
$(2N)^2$ constraints in a DBM at once.

In practice, we suggest to analyze integer variables in ${\mathbb{Q}}$ or 
${\mathbb{R}}$, as it is commonly done for polyhedron analysis \cite{poly}.
This method will add {\em noise} solutions, which is safe in the abstract
interpretation framework because we are only interested in an upper 
approximation of program behaviors.

\section{Operators and Transfer Functions}

In this section, we describe how to implement
the abstract operators and transfer functions needed for static analysis.

These are the generic ones described in \cite{interv} for the interval domain,
and in \cite{poly} for the polyhedron domain:
assignments, tests, control flow junctions and loops.
See Section VIII for an insight on how to use theses operators to 
actually build an analyzer.
If our abstract numerical domain is used in a more complex analysis 
or in a parameterized abstract domain
(backward and interprocedural analysis, such as in 
Bourdoncle's {\sc Syntox} analyzer, 
Deutsch's pointer analysis \cite{deutsch}, etc.),
one may need to add some more operators.

All the operators and transfer functions presented in this section obviously
respect coherence and are adapted from our PADO-II article \cite{pado2}.

\subsection{Equality and Inclusion Testing.}
We distinguish two cases. If one or both ${\mathcal{V}}^+$-domains are 
empty, then the test is obvious.
If none are empty, we use the following theorem
which relies on the properties of the strong closure:
\begin{mytheoreml}
\label{equalincluthm}
\item ${\mathcal{D}}^+(\vec{m}^+)\subseteq{\mathcal{D}}^+(\vec{n}^+)
\iff (\vec{m}^+)^\bullet\trianglelefteqslant\vec{n}^+$;
\item ${\mathcal{D}}^+(\vec{m}^+)={\mathcal{D}}^+(\vec{n}^+)
\iff (\vec{m}^+)^\bullet=(\vec{n}^+)^\bullet$.
\end{mytheoreml}

\subsection{Projection.}
Thanks to the saturation property of the strong closure, we can 
easily extract from a DBM $\vec{m}^+$ representing 
a non-empty octagon,
the interval in which a variable $v_i$ ranges~:

\begin{mytheorem}~\\
$\begin{array}{l}
\{\;t\;|\;\exists(s_0,\ldots,s_{N-1})\in{\mathcal{D}}^+(\vec{m}^+)
\mbox{ such that }s_i=t\;\}\\
\quad=[\;-(\vec{m}^+)^\bullet_{2i\;2i+1}/2,\;
(\vec{m}^+)^\bullet_{2i+1\;2i}/2\;]
\end{array}$

(interval bounds are included only if finite).
\end{mytheorem}

\subsection{Union and Intersection.}
The {\it max} and {\it min} 
operators on $\overline{{\mathbb{I}}}$ lead to point-wise
least upper bound $\vee$ and greatest lower bound $\wedge$ 
(with respect to the $\trianglelefteqslant$ order) operators on 
DBMs:
$$
\begin{array}{l}
\left[\vec{m}^+\wedge\vec{n}^+\right]_{ij}\deltaeq\min(\vec{m}^+_{ij},\vec{n}^+_{ij});\\
\left[\vec{m}^+\vee\vec{n}^+\right]_{ij}\deltaeq\max(\vec{m}^+_{ij},\vec{n}^+_{ij})
\enspace.
\end{array}
$$

These operators are useful to compute intersections and unions of octagons:
\begin{mytheoreml}
\label{unioninterthm}
\item ${\mathcal{D}}^+(\vec{m}^+\wedge\vec{n}^+)={\mathcal{D}}^+(\vec{m}^+)
\cap{\mathcal{D}}^+(\vec{n}^+)$.
\item ${\mathcal{D}}^+(\vec{m}^+\vee\vec{n}^+)\supseteq{\mathcal{D}}^+(\vec{m}^+)
\cup{\mathcal{D}}^+(\vec{n}^+)$.
\item If $\vec{m}^+$ and $\vec{n}^+$ represent non-empty octagons,
then:
\begin{center}
$\begin{array}{l}
((\vec{m}^+)^\bullet)\vee((\vec{n}^+)^\bullet)=\\
\quad\inf_{\trianglelefteqslant}\{\vec{o}^+\;|\;
{\mathcal{D}}^+(\vec{o}^+)\supseteq{\mathcal{D}}^+(\vec{m}^+)
\cup{\mathcal{D}}^+(\vec{n}^+)\}.
\end{array}$
\end{center}
\end{mytheoreml}

Remark that the intersection is always exact, but the union of
two octagons is not always an octagon, so we compute an upper approximation.
In order to get the best---smallest---approximation for the union, we
need to use the strong closure algorithm, as
stated in Theorem  \ref{unioninterthm}.3.

Another consequence of Theorem  \ref{unioninterthm}.3 is that if the two 
arguments 
of $\vee$ are strongly closed, then the result is also strongly closed.
Dually, the arguments of $\wedge$ do not need to be strongly closed in order
to get the best precision, but the result is seldom strongly closed---even if
the arguments are.
This situation is similar to what is described in our PADO-II
article  \cite{pado2}. Shaham, Kolodner, and Sagiv fail to
analyze this result in their CC2000 article \cite{mooly} and 
perform a useless closure after the union operator.

\subsection{Widening.}
Program semantics often use {\em fixpoints} to model arbitrary long 
computations such as {\em loops}.
Fixpoints are not computable in the octagon domain---as it is often the case
for abstract domains---because it is of infinite height.
Thus, we define a {\em widening operator}, as introduced in P.Cousot's thesis
\cite[\S 4.1.2.0.4]{these}, to
compute iteratively an upper approximation of the 
least fixpoint $\bigvee_{i\in\mathbb{N}}F^i(\vec{m}^+)$ 
greater than $\vec{m}^+$ of an operator $F$:
$$
\left[\vec{m}^+\triangledown\vec{n}^+\right]_{ij}\deltaeq
\left\{\begin{array}{ll}
\vec{m}^+_{ij}\quad&\mbox{ if }\vec{n}^+_{ij}\leq\vec{m}^+_{ij},\\
+\infty&\mbox{elsewhere}\enspace.
\end{array}\right.
$$

The idea behind this widening is to remove in $\vec{m}^+$ the constraints
that are not {\em stable} by union with $\vec{n}^+$; thus it is very similar
to the standard widenings used on the domains of intervals \cite{interv} 
and polyhedra \cite{poly}.
\cite{mooly} proposes a similar widening on the set of DBMs representing
${\mathcal{V}}$-domains.

The following theorem proves that $\triangledown$ is a widening in the 
octagon domain:
\begin{mytheoreml}
\label{wideningthm}
\item ${\mathcal{D}}^+(\vec{m}^+\triangledown\vec{n}^+)\supseteq
{\mathcal{D}}^+(\vec{m}^+)\cup{\mathcal{D}}^+(\vec{n}^+)$.
\item For all chains $(\vec{n}^+_i)_{i\in\mathbb{N}}$, the chain defined by
induction:
$$
\vec{m}^+_i\deltaeq\left\{
\begin{array}{ll}
(\vec{n}^+_0)^\bullet&\mbox{ if }i=0,\\
\vec{m}^+_{i-1}\triangledown((\vec{n}^+_i)^\bullet)\quad&\mbox{ elsewhere},\\
\end{array}
\right.
$$
is increasing, ultimately stationary, and with a limit
greater than $\bigvee_{i\in\mathbb{N}}(\vec{n}^+_i)^\bullet$.
\end{mytheoreml}

As for the union operator, the precision of the $\triangledown$ operator is
improved if its right argument is strongly closed; this is why we ensure the 
strong closure of $\vec{n}^+_i$ when computing $\vec{m}^+_i$ in 
Theorem  \ref{wideningthm}.2.

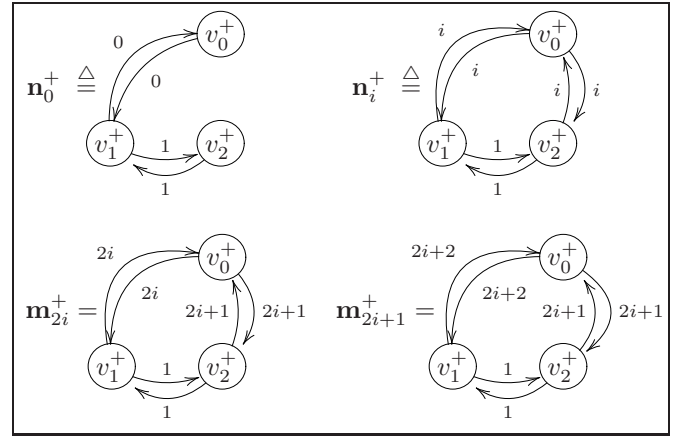
\begin{figure}
\begin{center}
\fbox{
\begin{tabular}{cc}
\hspace*{-0.4cm}
$\vec{n}^+_0\deltaeq$
\hspace*{-0.6cm}
\raisebox{0.8cm}{
\xymatrix{& 
*+[o][F-]{v^+_0} \ar@/_1pc/[dl]^{0} & \\
*+[o][F-]{v^+_1} \ar@/^1.5pc/[ur]^{0} \ar@/_0.5pc/[r]^{1} &
*+[o][F-]{v^+_2} \ar@/^1.0pc/[l]^{1} }}
&
\hspace*{-0.4cm}
$\vec{n}^+_i\deltaeq$
\hspace*{-0.5cm}
\raisebox{0.8cm}{
\xymatrix{& 
*+[o][F-]{v^+_0} \ar@/_1.5pc/[dl]^{i} \ar@/^1pc/[d]^{i} & \\
*+[o][F-]{v^+_1} \ar@/^2pc/[ur]^{i}   \ar@/_0.5pc/[r]^{1} & 
*+[o][F-]{v^+_2} \ar@/_0.5pc/[u]^{i}  \ar@/^1.0pc/[l]^{1} }}
\hspace*{-0.6cm}
\\ \\
\hspace*{-0.4cm}
$\vec{m}^+_{2i}=$
\hspace*{-0.5cm}
\raisebox{0.8cm}{
\xymatrix{& 
*+[o][F-]{v^+_0} \ar@/_1.5pc/[dl]^{2i}  \ar@/^1.0pc/[d]^{2i+1} & \\
*+[o][F-]{v^+_1} \ar@/^2pc/[ur]^{2i}    \ar@/_0.5pc/[r]^{1} & 
*+[o][F-]{v^+_2} \ar@/_0.5pc/[u]^{2i+1} \ar@/^1.0pc/[l]^{1} }}
&
\hspace*{-0.4cm}
$\vec{m}^+_{2i+1}=$
\hspace*{-0.8cm}
\raisebox{0.8cm}{
\xymatrix{& 
*+[o][F-]{v^+_0} \ar@/_1.5pc/[dl]^{2i+2} \ar@/^1.5pc/[d]^{2i+1} & \\
*+[o][F-]{v^+_1} \ar@/^2pc/[ur]^{2i+2}   \ar@/_0.5pc/[r]^{1} &
*+[o][F-]{v^+_2} \ar@/_1pc/[u]^{2i+1}    \ar@/^1.0pc/[l]^{1} }}
\hspace*{-0.6cm}
\end{tabular}
}
\end{center}
\caption{Example of an infinite strictly increasing chain defined by
$\vec{m}^+_0=(\vec{n}^+_0)^{\bullet},\;
\vec{m}^+_i=(\vec{m}^+_{i-1}\triangledown((\vec{n}^+_i)^\bullet)^\bullet$.
Remark that the nodes $\{v^-_0,v^-_1,v^-_2\}$ are not represented here due to
lack of space; this part of the DBMs can be easily figured out
by coherence.}
\label{wideningfig}
\end{figure}

One can be tempted to force the strong closure of the left argument of the
widening by replacing the induction step in Theorem \ref{wideningthm}.2 by:
$\vec{m}^+_i=(\vec{m}^+_{i-1}
\triangledown((\vec{n}^+_i)^\bullet))^\bullet\mbox{ if }i>0$.
However, we cannot do this safely as Theorem \ref{wideningthm}.2 is no
longer valid: one can build a strictly increasing infinite chain
$(\vec{m}^+_i)_{i\in{\mathbb{N}}}$ (see Figure \ref{wideningfig})
which means that fixpoints using this induction may not be computable!
This situation is similar to what is described in our PADO-II
article \cite{pado2}. Shaham, Kolodner, and Sagiv fail to
analyze this problem in their CC2000 article \cite{mooly}
and pretend all their computation are performed with closed DBMs.
If we want our analysis to terminate, it is very important {\it not to 
close} the $(\vec{m}^+_i)_{i\in{\mathbb{N}}}$ in the induction computation.

%\subsection{Narrowing.}
%\mydef{Narrowing operators} were introduced by P.Cousot
%in his thesis \cite[\S 4.1.2.0.11]{these}
%in order to try and recover, in a finite time,
%as much information as possible after some crude widening applications.
%Our narrowing is defined by:
%$$\left[\vec{m}^+\triangle \vec{n}^+\right]_{ij}\deltaeq
%\left\{
%\begin{array}{ll}
%\vec{n}^+_{ij} \quad& \mbox{if }\vec{m}^+_{ij}=+\infty,\\
%\vec{m}^+_{ij} \quad& \mbox{elsewhere}\enspace.
%\end{array}
%\right.
%$$
%
%The following theorem proves that $\triangle$ is a narrowing:
%\begin{mytheoreml}
%\label{narrowingthm}
%\item If ${\mathcal{D}}^+(\vec{n}^+)\subseteq{\mathcal{D}}^+(\vec{m}^+)$,\\
%then 
%${\mathcal{D}}^+(\vec{n}^+)
%\subseteq{\mathcal{D}}^+(\vec{m}^+\triangle\vec{n}^+)\subseteq
%{\mathcal{D}}^+(\vec{m}^+)$.
%\item  For all chains $(\vec{n}_i^+)_{i\in\mathbb{N}}$
%decreasing for $\trianglelefteqslant$, the chain defined by:
%$$
%\vec{m}^+_i\deltaeq
%\left\{
%\begin{array}{ll}
%(\vec{n}^+_0)^\bullet&\mbox{ if }i=0,\\
%((\vec{m}^+_{i-1})^\bullet) \triangle ((\vec{n}^+_i)^\bullet)
%\quad&\mbox{ elsewhere},
%\end{array}
%\right.
%$$
%is decreasing, ultimately stationary, and with a limit greater than
%$\bigwedge_{i\in\mathbb{N}}(\vec{n}^+_i)^\bullet$.
%\end{mytheoreml}

\subsection{Guard and Assignment.}
In order to analyze programs, we need to model the effect of {\em tests}
and {\em assignments}.

Given a DBM $\vec{m}^+$ that represents a set of 
possible values of the variables ${\mathcal{V}}^+$ at a program point, 
an arithmetic comparison $g$, a variable
$v_i\in{\mathcal{V}}^+$, and an arithmetic expression  $e$, 
we denote by $\vec{m}^+_{(g)}$ and $\vec{m}^+_{(v_i\leftarrow e)}$
DBMs representing respectively the set of possible values of ${\mathcal{V}}^+$
if the test $g$ succeeds and after the assignment 
$v_i\leftarrow e(v_0,\ldots,v_{N-1})$.
Since the exact representation of the resulting set is, in general, impossible,
we will only try to compute an upper approximation:

\begin{mypropertyl}
\label{guardassignthm}
\item ${\mathcal{D}}^+(\vec{m}^+_{(g)})\supseteq
\{s\in{\mathcal{D}}^+(\vec{m}^+)\;|\;s\mbox{ satisfies }g\}$.
\item ${\mathcal{D}}^+(\vec{m}^+_{(v_i\leftarrow e)})\supseteq
\{s[s_i\leftarrow e(s)]\;|\;s\in{\mathcal{D}}^+(\vec{m}^+)\}$\\
(where $s[s_i\leftarrow x]$ means $s$ with its $i^{\mbox{th}}$ component
changed into $x$).
\end{mypropertyl}

Here is an example definition:
\begin{mydefinitionl}
\label{guardassigndef}
\item $\left[\vec{m}^+_{(v_k+v_l\leq c)}\right]_{ij}\deltaeq$\nopagebreak

\hspace*{0.3cm}
$\begin{array}{l}
\left\{\begin{array}{ll}
\min(\vec{m}^+_{ij},c)&\mbox{if }(j,i)\in\{(2k,2l+1);(2l,2k+1)\},\\
\vec{m}^+_{ij}&\mbox{elsewhere},
\end{array}\right.
\end{array}$\nopagebreak

and similarly for $\vec{m}^+_{(v_k-v_l\leq c)}$
and $\vec{m}^+_{(-v_k-v_l\leq c)}\enspace.$
\bigskip

\item $\vec{m}^+_{(v_k\leq c)}\deltaeq\vec{m}^+_{(v_k+v_k\leq 2c)}$, and\nopagebreak

\hspace*{1.3em}$\vec{m}^+_{(v_k\geq c)}\deltaeq\vec{m}^+_{(-v_k-v_k\leq -2c)}\enspace.$
\bigskip

\item $\vec{m}^+_{(v_k+v_l=c)}\deltaeq
(\vec{m}^+_{(v_k+v_l\leq c)})_{(-v_k-v_l\leq -c)}$,\\
and similarly for $\vec{m}^+_{(v_k-v_l=c)}\enspace.$
\bigskip

\item
$\left[\vec{m}^+_{(v_k\leftarrow v_k+c)}\right]_{ij}\deltaeq
\vec{m}^+_{ij}+(\alpha_{ij}+\beta_{ij}) c$,
with\nopagebreak

\quad\quad$\alpha_{ij}\deltaeq
\left\{\begin{array}{ll}
+1\quad&\mbox{ if }j=2k,\\
-1&\mbox{ if }j=2k+1,\\
0&\mbox{ elsewhere },
\end{array}\right.$ 

and\nopagebreak

\quad\quad$\beta_{ij}\deltaeq
\left\{\begin{array}{ll}
-1\quad&\mbox{ if }i=2k,\\
+1&\mbox{ if }i=2k+1,\\
0&\mbox{ elsewhere }\enspace.
\end{array}\right.$
\bigskip

\item
$\left[\vec{m}^+_{(v_k\leftarrow v_l+c)}\right]_{ij}\deltaeq$\nopagebreak

\quad\quad$\left\{\begin{array}{ll}
c\quad&\mbox{ if }(j,i)\in\{(2k,2l);(2l+1,2k+1)\},\\
-c\quad&\mbox{ if }(j,i)\in\{(2l,2k);(2k+1,2l+1)\},\\
(\vec{m}^+)^\bullet_{ij}\quad&\mbox{ if }i,j\notin\{2k,2k+1\},\\
+\infty&\mbox{ elsewhere},
\end{array}\right.$\nopagebreak

for $k\neq l$.
\bigskip

\item In all other cases, we simply choose:

$\begin{array}{l}
\vec{m}^+_{(g)}\deltaeq\vec{m}^+,\\
\left[\vec{m}^+_{(v_k\leftarrow e)}\right]_{ij}\deltaeq
\left\{\begin{array}{ll}
(\vec{m}^+)^\bullet_{ij}&\mbox{ if }i,j\notin\{2k,2k+1\},\\
+\infty&\mbox{elsewhere}\enspace.
\end{array}\right.
\end{array}$
\end{mydefinitionl}

Remark that the assignment destroys informations about $v_k$ 
and this could result in some implicit constraints about other variables being
destroyed as well. To avoid precision degradation, we use
constraints from the strongly closed form $(\vec{m}^+)^\bullet_{ij}$ in 
Definitions \ref{guardassigndef}.5 and \ref{guardassigndef}.6.

Remark also that the guard and assignment transfer functions are exact,
except in the last---general---case of Definition  \ref{guardassigndef}.
There exists certainly many ways to improve the precision of Definition 
\ref{guardassigndef}.6.
For example, in order to handle arbitrary assignment $v_k\leftarrow e$,
one can use the projection operator to extract the interval
where the variables range, then use a simple interval arithmetic to compute
an approximation interval $[-e^-/2,e^+/2]$ where ranges the result
$$
\begin{array}{ll}
[-e^-,e^+]\supseteq e(&
[-(\vec{m}^+)^\bullet_{01},(\vec{m}^+)^\bullet_{10}],\;\ldots\;,\\
&[-(\vec{m}^+)^\bullet_{2N-2\;2N-1},(\vec{m}^+)^\bullet_{2N-1\;2N-2}]\;)
\end{array}
$$
and put back this information into $\vec{m}^+$:
$$
\left[\vec{m}^+_{(v_k\leftarrow e)}\right]_{ij}\deltaeq
\left\{\begin{array}{ll}
(\vec{m}^+)^\bullet_{ij}\quad&
\mbox{ if }i,j\notin\{2k,2k+1\},\\
e^+&\mbox{ if }(i,j)=(2k+1,2k),\\
e^-&\mbox{ if }(i,j)=(2k,2k+1),\\
+\infty&\mbox{ elsewhere}\enspace.
\end{array}\right.
$$

Finally, remark that we can extend easily the guard operator to {\it boolean 
formulas} with the following definition:
\begin{mydefinitionl}
\item $\vec{m}^+_{(g_1\;{\bf and}\;g_2)}\deltaeq
\vec{m}^+_{(g_1)}\wedge\vec{m}^+_{(g_2)}$;
\item $\vec{m}^+_{(g_1\;{\bf or}\;g_2)}\deltaeq
((\vec{m}^+_{(g_1)})^\bullet)\vee((\vec{m}^+_{(g_2)})^\bullet)$;
\item $\vec{m}^+_{(\neg g1)}$ is settled by the classical 
transformation:

\quad\quad$\begin{array}{lcl}
\neg (g_1\;{\bf and}\;g_2)&\rightarrow&(\neg g_1)\;{\bf or}\;(\neg g_2)\\
\neg (g_1\;{\bf or}\;g_2)&\rightarrow&(\neg g_1)\;{\bf and}\;(\neg g_2)
\enspace.
\end{array}$
\end{mydefinitionl}

\section{Lattice Structures}

In this section, we design two lattice structures: one on the set of
coherent  DBMs and one on the set of strongly closed DBMs.
The first one is useful to analyze fixpoint transfers between
abstract and concrete semantics, and the second one
allows us to design a meaning function---or 
even a Galois connection---linking the set of
octagons to the concrete lattice
${\mathcal{P}}({\mathcal{V}}^+\mapsto{\mathbb{I}})$, following the
abstract interpretation framework
described in Cousot and Cousot's POPL'79 article \cite{semdesign}.

Lattice structures and Galois connections can be used to
simplify proofs of correctness of static analyses---see, for example,
the author's Master thesis \cite{dea} for a proof of the correctness
of the analysis described in Section VIII.

\subsection{Coherent DBMs Lattice.}
The set ${\mathcal{M}}^+$ of coherent DBMs, 
together with the \mydef{order relation} $\trianglelefteqslant$
and the point-wise \mydef{least upper bound} $\vee$ and 
\mydef{greatest lower bound}
$\wedge$, is almost a lattice. 
It only needs a \mydef{least element $\bot$}, so we
extend $\trianglelefteqslant$, $\vee$ and $\wedge$ to
${\mathcal{M}}^+_{\bot}={\mathcal{M}}^+\cup\{\bot\}$ in an obvious way to get
$\sqsubseteq$, $\sqcup$ and $\sqcap$.
The \mydef{greatest element $\top$} is the DBM with all its coefficients 
equal to $+\infty$.

\begin{mytheoreml}
\item $({\mathcal{M}}^+_{\bot},\sqsubseteq,\sqcap,\sqcup,\bot,\top)$ is a lattice.
\item This lattice is complete if $({\mathbb{I}},\leq)$ is complete
(${\mathbb{I}}={\mathbb{Z}}$ or ${\mathbb{R}}$, but not ${\mathbb{Q}}$).
\end{mytheoreml}

There are, however, two problems with this lattice.
First, this lattice is not isomorphic to a sub-lattice of
${\mathcal{P}}({\mathcal{V}}^+\mapsto\mathbb{I})$ as two different
DBMs can have the same ${\mathcal{V}}^+$-domain.
Then, the least upper bound operator $\sqcup$ is not the most
precise upper approximation of the union of 
two octagons because we do not force the arguments to be strongly closed.

\subsection{Strongly Closed DBMs Lattice.}
To overcome these difficulties, we build another lattice, based
on strongly closed DBMs.
First, consider the set ${\mathcal{M}}^{\bullet}_{\bot}$ of strongly closed 
DBMs
${\mathcal{M}}^{\bullet}$, with a \mydef{least element $\bot^{\bullet}$} added.
Now, we define a \mydef{greatest element $\top^{\bullet}$}, 
a \mydef{partial order relation $\sqsubseteq^{\bullet}$},
a \mydef{least upper bound $\sqcup^{\bullet}$} and 
a \mydef{greatest lower bound $\sqcap^{\bullet}$} in ${\mathcal{M}}^{\bullet}_{\bot}$ as follows:
\medskip

$
\vec{\top^{\bullet}}_{ij}\deltaeq
\left\{
\begin{array}{ll}
0&\mbox{if }i=j,\\
+\infty\quad&\mbox{elsewhere},
\end{array}\right.
$
\medskip

$
\vec{m}^+\sqsubseteq^{\bullet}\vec{n}^+\deltaiff
\left\{
\begin{array}{ll}
\mbox{either}&\vec{m}^+=\bot^{\bullet},\\
\mbox{or}&\vec{m}^+,\vec{n}^+\neq \bot^{\bullet},\;
\vec{m}^+\trianglelefteqslant\vec{n}^+,
\end{array}\right.
$
\medskip

$
\vec{m}^+\sqcup^{\bullet}\vec{n}^+\deltaeq
\left\{
\begin{array}{ll}
\vec{m}^+&\mbox{if }\vec{n}^+=\bot^{\bullet},\\
\vec{n}^+&\mbox{if }\vec{m}^+=\bot^{\bullet},\\
\vec{m}^+\vee\vec{n}^+\quad&\mbox{elsewhere},
\end{array}\right.
$
\medskip

$
\vec{m}^+\sqcap^{\bullet}\vec{n}^+\deltaeq
\left\{
\begin{array}{ll}
\vec{\bot^{\bullet}}&\mbox{if }\bot^{\bullet}\in\{\vec{m}^+,\vec{n}^+\}
\mbox{ or}\\
&\;\;{\mathcal{D}}^+(\vec{m}^+\wedge\vec{n}^+)=\emptyset,\\
(\vec{m}^+\wedge\vec{n}^+)^{\bullet}\quad&\mbox{elsewhere}\enspace.
\end{array}\right.
$
\medskip

Thanks to Theorem \ref{equalincluthm}.2,
every non-empty octagon has a unique 
representation in ${\mathcal{M}}^{\bullet}$; $\bot^{\bullet}$ is the representation for
the empty set.
We build a \mydef{meaning function $\gamma$} which is an extension
of $\cdot\mapsto{\mathcal{D}}^+(\cdot)$ to ${\mathcal{M}}^{\bullet}_{\bot}$:
$$
\gamma(\vec{m}^+)\deltaeq
\left\{
\begin{array}{ll}
\emptyset&\mbox{if }\vec{m}^+=\bot^{\bullet},\\
{\mathcal{D}}^+(\vec{m}^+)\quad&\mbox{elsewhere}\enspace.
\end{array}\right.
$$

\begin{mytheoreml}
\item $({\mathcal{M}}^{\bullet}_{\bot},\sqsubseteq^{\bullet},\sqcap^{\bullet},\sqcup^{\bullet},\bot^{\bullet},\top^{\bullet})$ 
is a lattice and $\gamma$ is one-to-one.
\item If $({\mathbb{I}},\leq)$ is complete, this lattice is complete and
$\gamma$ is meet-preserving: $\gamma(\bigsqcap^{\bullet}X)=
\bigcap\{\gamma(x)\;|\;x\in X\}$. 
We can---according to Cousot and Cousot \cite[Prop. 7]{fixpoint}---build 
a canonical \mydef{Galois insertion}:
$${\mathcal{P}}({\mathcal{V}}^+\mapsto{\mathbb{I}})\;
\galoiS{\alpha}{\gamma}\;
{\mathcal{M}}^{\bullet}_{\bot}$$
where the \mydef{abstraction function $\alpha$} is defined by:\\
$\alpha(X)=\bigsqcap^{\bullet}\;\{\;
x\in{\mathcal{M}}^{\bullet}_{\bot}\;|\;X\subseteq\gamma(x)\;\}\enspace.$
\end{mytheoreml}

The ${\mathcal{M}}^{\bullet}_{\bot}$ lattice features a nice meaning function
and a precise union approximation;
thus, it is tempting to force all our operators and transfer functions to 
live in ${\mathcal{M}}^{\bullet}_{\bot}$ by forcing strong 
closure on their result.
However, we saw this does not work for the widening, so fixpoint computations
{\em must} be performed in the ${\mathcal{M}}^+_{\bot}$ lattice.

\section{Application to Program Analysis}

In this section, we present the program analysis based on our new domain
that enabled us to prove the correctness of the program in Figure
\ref{sampleprog}.

This is only one example application of our domain
for program analysis purpose.
It was chosen for its simplicity of presentation and implementation.
A fully featured tool that can deal with real-life programs, taking care of
pointers, procedures and objects is far beyond the scope of this work.
However, current tools using the interval or the polyhedron domains
could benefit from this new abstract domain.

\subsection{Presentation of the Analysis.}

Our analyzer is very similar to the one described in Cousot and Halbwachs's 
POPL'78 article \cite{poly}, except
it uses our new abstract domain instead of the abstract domain of polyhedra.

Here is a sketched description of this analysis---more informations, 
as well as proofs of its correctness can be found in
the author's Master thesis \cite{dea}.

We suppose that our program is procedure-free, has only numerical
variables---no pointers or array---and is solely composed of assignments, 
{\bf if then else fi} and {\bf while do done} statements.
Syntactic program locations $l_i$ are placed to visualize the control flow:
there are locations before and after statements, at the beginning and the end
of {\bf then} and {\bf else} branches and inner loop blocks; the location
at the program entry point is denoted by $l_0$.

The analyzer associates to each program point $l_i$ an element
$\vec{m}^+_i\in{\mathcal{M}}^+_\bot$.
At the beginning, all $\vec{m}^+_i$ are $\bot$ (meaning the control flow
cannot pass there) except $\vec{m}^+_0=\top$.
Then, informations are propagated through the control flow as if the program
were executed:
\begin{itemize}
\item For $\llbracket(l_i)\;v_i\leftarrow e\;(l_{i+1})\rrbracket$,
we set $\vec{m}^+_{i+1}=(\vec{m}^+_i)_{(v_i\leftarrow e)}$.
\item For a test $\llbracket(l_i)\;{\bf if}\;g\;{\bf then}\;(l_{i+1})\;\cdots\;
{\bf else}\;(l_j)\;\cdots\rrbracket$, 
we set $\vec{m}^+_{i+1}=(\vec{m}^+_i)_{(g)}$ and
$\vec{m}^+_{j}=(\vec{m}^+_i)_{(\neg g)}$.
\item When the control flow merges after a test
$\llbracket{\bf then}\;\cdots\;
(l_i)$ ${\bf else}\;\cdots\;(l_j)\;{\bf fi}\;(l_{j+1})\rrbracket$,
we set $\vec{m}^+_{j+1}=((\vec{m}^+_i)^\bullet)\sqcup
((\vec{m}^+_j)^\bullet)$.
\item For a loop $\llbracket\;(l_i)\;{\bf while}\;g\;{\bf do}\;(l_j)\cdots\;(l_k)\;
{\bf done}\;(l_{k+1})\rrbracket$, we must solve the relation
$\vec{m}^+_j=(\vec{m}^+_i\sqcup\vec{m}^+_k)_{(g)}$.
We solve it iteratively using the widening:
suppose $\vec{m}^+_i$ is known and we can deduce a $\vec{m}^+_k$ from
any $\vec{m}^+_j$ by propagation; we compute the limit $\vec{m}^+_{j}$ of
$$
\left\{\begin{array}{l}
\vec{m}^+_{j,0}=(\vec{m}^+_i)_{(g)}\\
\vec{m}^+_{j,n+1}=\vec{m}^+_{j,n}\triangledown 
  ((\vec{m}^+_{k,n})^\bullet_{(g)})
\end{array}\right.
$$
then $\vec{m}^+_k$ is computed by propagation of $\vec{m}^+_j$ and
we set $\vec{m}^+_{k+1}=((\vec{m}^+_i)^\bullet_{(\neg g)})\sqcup
((\vec{m}^+_k)^\bullet_{(\neg g)})$
\end{itemize}

At the end of this process, each $\vec{m}^+_i$ is a valid invariant
that holds at program location $l_i$.
This method is called {\it abstract execution}.

\subsection{Practical Results.}

The analysis described above has been implemented in {\sf OCaml} and used
on a small set of rather simple algorithms.

Figure \ref{detailedprog} shows the detailed computation for the lines
5--9 from Figure \ref{sampleprog}.
Remark that the program has been adapted to the language described in the
previous section, and program locations $l_0$,\ldots,$l_9$ have been added.
Also, for the sake of brevity, DBMs are presented in equivalent constraint set 
form, and only the useful constraints are shown.
Thanks to the widening, the fixpoint is reached after only two iterations:
invariants $\vec{m}^+_{k,0,\;k=2\ldots 8}$ only hold in the {\it first} 
iteration of the loop ($i=1$); invariants $\vec{m}^+_{k,1,\;k=2\ldots 8}$
hold for {\it all} loop iterations $(1\leq i \leq m)$.
At the end of the analysis, we have $(-m\leq a\leq m)\in(\vec{m}^+_9)^\bullet$.

Our analyzer was also able to prove that the well-known 
Bubble sort and Heap sort do
not perform out-of-bound error while accessing array elements
and to prove that Lamport's Bakery algorithm \cite{bakery}
for synchronizing two processes is correct---however, 
unlike the example in Figure \ref{sampleprog}, these analysis
where already in the range of our PADO-II article \cite{pado2}.

\begin{figure}
\begin{center}
\fbox{
\begin{tabular}{l}
\hspace*{-0.5cm}
%\textsf{
\sffamily
\begin{tabular}{rl}
4&$(l_0)$ $a\leftarrow 0$; $i\leftarrow 1$ $(l_1)$\\
&{\bf while} $i\leq m$ {\bf do} $(l_2)$\\
7&\quad {\bf if} $?$\\
8&\quad\quad {\bf then} $(l_3)$ $a\leftarrow a+1$ $(l_4)$\\
9&\quad\quad {\bf else} $(l_5)$ $a\leftarrow a-1$ $(l_6)$\\
&\quad {\bf fi} $(l_7)$\\
&\quad $i\leftarrow i+1$ $(l_8)$\\
11&{\bf done} $(l_9)$
\end{tabular}
%}
\normalfont
\\ \hline 
\hspace*{-0.6cm}
$\begin{array}{l}
\vec{m}^+_0=\top \\
\vec{m}^+_1=\{i=1;\;a=0;\;1-i\leq a\leq i-1\}\\
\\
\mbox{\it First iteration of the loop}\\
\vec{m}^+_{2.0}=\{i=1;\;a=0;\;1-i\leq a\leq i-1;\;i\leq m\}\\
\vec{m}^+_{3,0}=\vec{m}^+_{5,0}=\vec{m}^+_{2.0}\\
\vec{m}^+_{4,0}=\{i=1;\;a=1;\;2-i\leq a\leq i;\;i\leq m\}\\
\vec{m}^+_{6,0}=\{i=1;\;a=-1;\;-i\leq a\leq i-2;\;i\leq m\}\\
\vec{m}^+_{7,0}=\{i=1;\;a\in[-1,1];\;-i\leq a\leq i;\;i\leq m\}\\
\vec{m}^+_{8,0}=\{i=2;\;a\in[-1,1];\;1-i\leq a\leq i-1;\;i\leq m+1\}\\
\\
\mbox{\it Second iteration of the loop}\\
\vec{m}^+_{2,1}=\vec{m}^+_{3,1}=\vec{m}^+_{5,1}=\vec{m}^+_{2,0}\;
\triangledown\;(\vec{m}^+_{8,0})_{(i\leq m)}\\
\hspace*{0.825cm}=\{1\leq i\leq m;\;1-i\leq a\leq i-1\}\\
\vec{m}^+_{4,1}=\{1\leq i\leq m;\;2-i\leq a\leq i\}\\
\vec{m}^+_{6,1}=\{1\leq i\leq m;\;-i\leq a\leq i-2\}\\
\vec{m}^+_{7,1}=\{1\leq i\leq m;\;-i\leq a\leq i\}\\
\vec{m}^+_{8,1}=\{2\leq i\leq m+1;\;1-i\leq a\leq i-1\}\\
\\
\mbox{\it Third iteration of the loop}\\
\vec{m}^+_{2,2}=\vec{m}^+_{2,1}\quad\quad\mbox{\it (fixpoint reached)}\\
\\
\vec{m}^+_2=\vec{m}^+_{2,1}\quad\quad
\vec{m}^+_8=\vec{m}^+_{8,1}\\
\vec{m}^+_9=\{i=m+1;\;1-i\leq a\leq i-1\}

\end{array}$
\hspace*{-0.7cm}
\end{tabular}
}
\end{center}
\caption{Detailed analysis of lines 5--9 from Figure \ref{sampleprog}.
For sake of conciseness DBMs are shown in their equivalent constraint
set form and useless constraints are not shown.}
\label{detailedprog}
\end{figure}

\subsection{Precision and Cost.}

The computation speed in our abstract domain is limited by the cost
of the strong closure algorithm because it is the most used and 
the most costly algorithm. Thus, most abstract operators have
a ${\mathcal{O}}(N^3)$ worst case time cost.
Because a fully featured tool using our domain is not yet available,
we do not know how well this analysis scales up to large programs.

The invariants computed are {\em always} more precise than 
the ones computed in \cite{pado2}, which gives itself always better results
than the widespread intervals domain \cite{interv}; but they are
less precise than the costly polyhedron analysis \cite{poly}.
Possible loss of precision have three causes: non-exact union, non-exact
guard and assignment transfer functions, and widening in loops.
The first two causes can be worked out by refining 
Definition  \ref{guardassigndef} and choosing to represent, as abstract state,
any finite union of octagons instead of a single one.
Promising representations are the
{\em Clock-Difference Diagrams}
(introduced in 1999 by Larsen, Weise, Yi, and Pearson \cite{CDD}) 
and {\em Difference Decision Diagrams} 
(introduced in M{\o}ller, Lichtenberg, Andersen, and Hulgaard's
CSL'99 paper \cite{DDD}), which are
tree-based structures introduced by the model-checking community
to efficiently represent finite unions of ${\mathcal{V}}^0$-domains, but they
need adaptation in order to be used in the abstract interpretation framework
and must be extended to octagons.

\section{Conclusion}

In this article, we presented a new numerical abstract domain that extends,
without much performance degradation, 
the DBM-based abstract domain described in our PADO-II article \cite{pado2}.
This domain allows us to manipulate invariants of the form
$(\pm x \pm y \leq c)$ with a ${\mathcal{O}}(n^2)$ worst case memory cost
per abstract state and a ${\mathcal{O}}(n^3)$ worst case time cost per abstract
operation---where $n$ is the number of variables in the program.

We claim that our approach is fruitful since it allowed us to prove
automatically the correctness of some non-trivial algorithms, beyond the
scope of interval analysis, for a much smaller cost than polyhedron analysis.
However, our prototype implementation did not allow us to test our
domain on real-life programs and we still do not know if it will scale up.
It is the author's hope that this new domain will be integrated into
currently existing static analyzers as an alternative to the intervals and
polyhedra domains.

% Bibliography
%%%%%%%%%%%%%%

\bibliographystyle{IEEE}
\bibliography{bibarticle}

\end{document}

%% file: vdomain.pstex_t
\begin{picture}(0,0)%
\epsfig{file=vdomain.pstex}%
\end{picture}%
\setlength{\unitlength}{1579sp}%
\begingroup\makeatletter\ifx\SetFigFont\undefined%
\gdef\SetFigFont#1#2#3#4#5{%
  \reset@font\fontsize{#1}{#2pt}%
  \fontfamily{#3}\fontseries{#4}\fontshape{#5}%
  \selectfont}%
\fi\endgroup%
\begin{picture}(3825,3024)(1951,-5773)
\put(5776,-5536){\makebox(0,0)[lb]{\smash{\SetFigFont{6}{7.2}{\rmdefault}{\mddefault}{\updefault}\special{ps: gsave 0 0 0 setrgbcolor}$v_1$\special{ps: grestore}}}}
\put(2401,-3511){\makebox(0,0)[lb]{\smash{\SetFigFont{6}{7.2}{\rmdefault}{\mddefault}{\updefault}\special{ps: gsave 0 0 0 setrgbcolor}$v_2$\special{ps: grestore}}}}
\put(1951,-5161){\makebox(0,0)[lb]{\smash{\SetFigFont{6}{7.2}{\rmdefault}{\mddefault}{\updefault}\special{ps: gsave 0 0 0 setrgbcolor}$v_0$\special{ps: grestore}}}}
\end{picture}

%% file: emptyZ.pstex_t
\begin{picture}(0,0)%
\special{psfile=emptyZ.pstex}%
\end{picture}%
\setlength{\unitlength}{1973sp}%
\begingroup\makeatletter\ifx\SetFigFont\undefined%
\gdef\SetFigFont#1#2#3#4#5{%
  \reset@font\fontsize{#1}{#2pt}%
  \fontfamily{#3}\fontseries{#4}\fontshape{#5}%
  \selectfont}%
\fi\endgroup%
\begin{picture}(3750,2941)(1801,-5537)
\put(1801,-5461){\makebox(0,0)[lb]{\smash{\SetFigFont{9}{10.8}{\rmdefault}{\bfdefault}{\updefault}{\color[rgb]{0,0,0}$v_1-v_0\leq 0$}%
}}}
\put(4051,-5461){\makebox(0,0)[lb]{\smash{\SetFigFont{9}{10.8}{\rmdefault}{\bfdefault}{\updefault}{\color[rgb]{0,0,0}$v_0+v_1\leq 3$}%
}}}
\put(5551,-4936){\makebox(0,0)[lb]{\smash{\SetFigFont{9}{10.8}{\rmdefault}{\bfdefault}{\updefault}{\color[rgb]{0,0,0}$v_0$}%
}}}
\put(4651,-3961){\makebox(0,0)[lb]{\smash{\SetFigFont{9}{10.8}{\rmdefault}{\bfdefault}{\updefault}{\color[rgb]{0,0,0}$2v_1\geq 3$}%
}}}
\put(2326,-2836){\makebox(0,0)[lb]{\smash{\SetFigFont{9}{10.8}{\rmdefault}{\bfdefault}{\updefault}{\color[rgb]{0,0,0}$v_1$}%
}}}
\end{picture}